\documentclass[11pt,english,abstracton]{scrartcl}
\usepackage[T1]{fontenc}
\usepackage[latin1]{inputenc}
\usepackage{geometry}
\usepackage{amsmath}
\usepackage{amssymb}

\makeatletter

\providecommand{\tabularnewline}{\\}

 \usepackage{verbatim}

\usepackage{graphicx}
\usepackage{ae,aecompl}
\usepackage{cite}
\usepackage{lastpage}

\renewcommand{\textrm}{\text}

\usepackage{babel}
\makeatother
\begin{document}

\titlehead{\begin{flushright}TUM--HEP--594/05\\
 OSU--HEP--05--09\end{flushright}}

\title{Vacuum Energy from an Extra Dimension with UV/IR Connection}

\author{Florian Bauer$^{a}$%
\footnote{Email: \texttt{fbauer@ph.tum.de}%
}~ and Gerhart Seidl$^{b}$%
\footnote{Email: \texttt{gerhart.seidl@okstate.edu}%
}\\
\\
 \textit{\large $^{a}$Physik--Department, Technische Universität München,}\\
 \textit{\large James--Franck--Straße, D--85748 Garching, Germany}\\
 \textit{\large $^{b}$Department of Physics, Oklahoma State University,}\\
 \textit{\large Stillwater, OK 74075, USA}}

\date{~}

\maketitle
\begin{abstract}
We propose a lower limit on the size of a single discrete gravitational
extra dimension in the context of an effective field theory for massive
gravitons. The limit arises in this setup from the requirement that
the Casimir energy density of quantum fields is in agreement with
the observed dark energy density of the universe $\rho_{\textrm{obs}}\simeq10^{-47}\,\textrm{GeV}^{4}$.
The Casimir energy densities can be exponentially suppressed to an
almost arbitrarily small value by the masses of heavy bulk fields,
thereby allowing a tiny size of the extra dimension. This suppression
is only restricted by the strong coupling scale of the theory, which
is known to be related to the compactification scale via an UV/IR
connection for local gravitational theory spaces. We thus obtain a
lower limit on the size of the discrete gravitational extra dimension
in the range $(10^{12}\,\textrm{GeV})^{-1}\dots(10^{7}\,\textrm{GeV})^{-1}$,
while the strong coupling scale is by a factor $\sim10^{2}$ larger
than the compactification scale. We also comment on a possible cancelation
of the gravitational contribution to the quantum effective potential. 
\end{abstract}

\section{Introduction}

Recent observations suggest that the universe is currently in a phase
of accelerated expansion \cite{Riess:1998cb,Perlmutter:1998np,Spergel:2003cb,Tegmark:2003ud,Boughn:2004zm},
that is assumed to be driven by an energy form with negative pressure
called Dark Energy~(DE). The most famous candidate for DE is a positive
Cosmological Constant~(CC), which is equivalent to a positive vacuum
energy density. Although DE represents the dominant part (about 75\%)
of the total energy density of the universe, the observed value of
the CC is only of the order $\rho_{\textrm{obs}}\simeq10^{-47}\:{\textrm{GeV}}^{4}$,
which is extremely small compared to usual particle physics scales.
So far, no generally accepted solution has been given to the problem
of understanding such a tiny value of the CC, which is known as the
CC problem \cite{Weinberg:1988cp}.

It has been emphasized, that a nonzero CC arising from the Casimir
effect \cite{Casimir:1948dh,Bordag:2001qi} in Kaluza--Klein (KK)
theories \cite{Kaluza:1921tu} might be relevant for the dynamical
compactification of extra dimensions \cite{Appelquist:1982zs,Candelas:1983ae,CasimirAdS}.
In this scenario, the Casimir energies produced by the fluctuations
of gravitational and massless matter fields propagating in the internal
space, would yield a contribution to DE which depends on the size
of the extra dimensions. DE could therefore provide via the Casimir
effect a probe of the geometric infrared (IR) structure of the higher--dimensional
manifold. It would now be interesting to see, whether the Casimir
energies contributing to DE, might also be sensitive to the ultraviolet
(UV) details of the theory. In fact, distinct higher--dimensional
gauge theories that reproduce similar physics in the IR, can look
drastically different in the UV. This may be best appreciated by the
example of dimensional deconstruction \cite{Arkani-Hamed:2001ca,Hill:2000mu},
which yields a class of manifestly gauge--invariant and renormalizable
effective Lagrangians for KK modes and thus represents a possible
UV completion of higher--dimensional gauge theories.%
\footnote{For an early application of similar ideas, see Ref.~ \cite{Halpern:1975yj}.%
} In this type of models, one could only observe at high energies that
the physics of extra dimensions actually emerges dynamically in a
purely four--dimensional (4D) setting, which denotes a radical departure
from the usual treatment of higher--dimensional theories near their
UV cutoff. Recently, the idea of deconstruction has also been applied
to an effective field theory for massive gravitons \cite{Arkani-Hamed:2002sp,Arkani-Hamed:2003vb,Schwartz:2003vj},
which is defined in a {}``theory space'' \cite{Arkani-Hamed:2001ed}
containing {}``sites'' and {}``links''. This allows the construction
of discrete gravitational extra dimensions, that show qualitatively
new properties as compared to non--gravitational theory spaces \cite{Arkani-Hamed:2003vb,Schwartz:2003vj}.
A major feature of discrete gravitational extra dimensions is, that
they exhibit a strong coupling scale $\Lambda$ in the UV, which depends
via an {}``UV/IR connection'' on the size or IR length--scale of
the compactified extra dimension \cite{Arkani-Hamed:2003vb}. We will
therefore have to expect that a contribution to DE arising from the
Casimir effect in discrete gravitational extra dimensions could be
related to the UV structure of the theory in a non--trivial way.

In this paper, we consider a vacuum energy contribution to DE, which
is generated from the Casimir effect in a single discrete gravitational
extra dimension. For this purpose, we treat the gravitational theory
space as a flat background for quantum fields propagating in the latticized
five--dimensional (5D) bulk. In determining the Casimir energy densities
of the latticized bulk fields, we assume linearized gravity and truncate
the theory at the 1--loop level. Since these energy densities contribute
to the CC, they have to lie below the observed value~$\rho_{\textrm{obs}}\sim10^{-47}\,\textrm{GeV}^{4}$,
associated with the accelerated expansion of the universe. For massless
bulk fields%
\footnote{A scenario for obtaining the observed CC from a 5D Casimir effect
of massless bulk matter fields with a sub--mm extra dimension has
been proposed, \textit{e.g.}, in Ref.~\cite{Milton:2001np}. Current
Cavendish--type experiments, however, put already very stringent upper
bounds of the order $R\lesssim0.1\:\textrm{mm}$ on the possible size~$R$
of extra dimensions \cite{Hoyle:2004cw}. Upper bounds on $R$ from the Casimir
effect in the presence of universal extra dimensions are also given in Ref.~
\cite{Poppenhaeger:2003es}.}, the 4D Casimir energy density $\rho$ scales with the size (circumference)
$R$ of the extra dimension as $|\rho|\sim R^{-4}$, which would lead
to a lower bound~$R\gtrsim(10^{-3}\,\textrm{eV})^{-1}\sim0.1\,\textrm{mm}$.
A much smaller size $R$ becomes possible, if the bulk fields have
nonzero masses $M_{X}$, in which case the Casimir energies are exponentially
suppressed for $M_{X}\gg R^{-1}$. In the discrete gravitational extra
dimension, this suppression is only limited by the strong coupling
scale $\Lambda$ of the theory, since in a sensible effective field
theory, $M_{X}$ should be smaller than the UV cutoff $\Lambda$.
By virtue of the UV/IR connection in minimal discretizations, however,
the cutoff $\Lambda$ depends on $R$ and can be much lower than the
usual 4D Planck scale $M_{\textrm{Pl}}\simeq10^{19}\,\textrm{GeV}$.
As a consequence, we expect from the Casimir effect a smallest possible
value or lower limit on the size $R$, when $M_{X}$ can at most be
as large as the strong coupling scale $\Lambda$.

The paper is organized as follows: In Sec.~\ref{sec:discretedimensions},
we review the model for a single discrete gravitational extra dimension
and briefly discuss the strong coupling behavior as the origin of
the UV/IR connection. In Sec.~\ref{sec:latticizedmatter}, we include
scalar and fermionic lattice fields in the gravitational theory space.
Sec.~\ref{sec:matterenergy} represents the main part of this work,
where we first consider the vacuum energy of quantum fields on the
transverse lattice and then determine the suppression of the Casimir
energy density due to large bulk masses of the latticized matter fields.
Then, we employ the UV/IR--connection and the observational constraints
on the DE density to derive a lower limit on the size of the extra
dimension. Finally, in Sec.~\ref{sec:Conclusions}, we present our
summary and conclusions.

\section{Review of discrete gravitational extra dimensions}

\label{sec:discretedimensions}

\label{sec:gravity} Recently, Arkani--Hamed and Schwartz have applied
general techniques for implementing gravity in theory space \cite{Arkani-Hamed:2002sp}
to a model for a single discrete gravitational extra dimension \cite{Arkani-Hamed:2003vb}.
In this section, we briefly review this model for a discrete gravitational
extra dimension, which describes pure gravity in the latticized bulk.
In the next section, we then extend this setup to a model, that also
includes matter fields.

\begin{figure}
\begin{center}\includegraphics*[bb = 178 703 432 744]{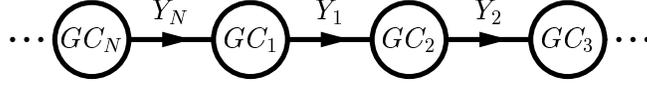}\end{center}

\vspace{2mm}

\caption{\label{fig:linear}Section of the gravitational theory space for
a discrete fifth dimension compactified on the circle $\mathcal{S}^{1}$.
Each site corresponds to one general coordinate invariance $GC_{i}$
($i=1,2,\ldots,N$), where two neighboring sites $i$ and $i+1$ are
connected by one link field $Y_{i}$ and we identify $i+N=i$.}
\end{figure}

Consider the minimal theory space for a single discrete gravitational
extra dimension proposed in Ref.~\cite{Arkani-Hamed:2003vb}, which
can be conveniently summarized by the {}``moose'' \cite{Georgi:1985hf}
or {}``quiver'' \cite{Douglas:1996sw} type diagram shown in Fig.~\ref{fig:linear}.%
 Each circle or site $i$, where $i=1,2,\ldots,N$, corresponds to
one general coordinate invariance (GC) symmetry $GC_{i}$ and is equipped
with a metric $g_{\mu\nu}^{i}$ for this site.%
\footnote{Multi--graviton theories have been considered earlier, \textit{e.g.},
in Ref.~\cite{Boulanger:2000rq} and in connection with discretized
brane--worlds in Ref.~\cite{Damour:2002wu}.%
} An arrow connecting two sites $i$ and $i+1$ symbolizes a link field
$Y_{i}$, which transforms as a vector under the two neighboring GC's.
Since we suppose for the sites the identification $i+N=i$, the theory
space is compactified on a circle. On each site $i$, we assume the
usual Einstein--Hilbert action, \textit{i.e.}, the purely gravitational
contribution from all sites to the total action is given by \begin{eqnarray}
\mathcal{S}_{\textrm{site}}^{g} & = & \sum_{i=1}^{N}\int\textrm{d}^{4}x\, M^{2}\sqrt{g^{i}}R(g^{i}),\label{eq:gsite}\end{eqnarray}
 where $R(g^{i})$ is the Ricci scalar on the site $i$, while~$M^{2}=M_{4}^{2}/N$
and $M_{4}=1/\sqrt{16\pi G_{\textrm{N}}}$ with $G_{\textrm{N}}$
as the 4D Newton's constant. We see in Eq.~(\ref{eq:gsite}), that
the action $\mathcal{S}_{\textrm{site}}^{g}$ is invariant under the
large GC product group $\Pi_{i=1}^{N}GC_{i}$. This $N$--fold product
GC, however, is explicitly broken by the gravitational interactions
$\mathcal{S}_{\textrm{link}}^{g}$ between the sites. In a minimal
discretization with only nearest neighbor interactions, the action
$\mathcal{S}_{\textrm{link}}^{g}$ is found to be on a Fierz--Pauli
\cite{Fierz:1939ix} form%
\footnote{The Fierz--Pauli form for graviton mass terms ensures the absence
of ghosts in the spectrum. For a recent discussion of ghosts in massive
gravity, see Ref.~\cite{Creminelli:2005qk}.%
}\begin{eqnarray}
\mathcal{S}_{\textrm{link}}^{g} & = & \sum_{i=1}^{N}\int\textrm{d}^{4}x\sqrt{g^{i}}M^{2}m^{2}(g_{\mu\nu}^{i}-g_{\mu\nu}^{i+1})(g_{\alpha\beta}^{i}-g_{\alpha\beta}^{i+1})(g^{i\mu\nu}g^{i\mu\nu}-g^{i\mu\alpha}g^{i\nu\beta}),\label{eq:FP}\end{eqnarray}
 where the inverse mass~$m^{-1}$ of the heaviest graviton sets the
lattice spacing $a=m^{-1}$, \textit{i.e.}, the discrete extra
dimension has a size (circumference) $R=N/m$ such that the 5D Planck
scale is given by $M_{5}=(M_{4}^{2}/R)^{1/3}$, which defines the
usual UV cutoff of the 5D theory. The product group $\Pi_{i=1}^{N}GC_{i}$
is explicitly broken by the action in Eq.~(\ref{eq:FP}) to the diagonal
GC. When we now expand in the weak field limit the metrics about flat
space as $g_{\mu\nu}^{i}=\eta_{\mu\nu}+h_{\mu\nu}^{i}$, where $\eta_{\mu\nu}$
is the Minkowski space metric, the mass--terms of the gravitons can
be written as \begin{equation}
\mathcal{S}_{ij}^{\textrm{FP}}=\int\textrm{d}^{4}x\, M^{2}m^{2}(2\delta_{i,j}-\delta_{i,j+1}-\delta_{i,j-1})(h_{\mu\nu}^{i}h^{\mu\nu,j}-h_{\mu}^{\mu,i}h_{\nu}^{\nu,j}),\label{eq:massterms}\end{equation}
 leading to a graviton spectrum with mass--squares \begin{equation}
m_{n}^{2}=4m^{2}\sin^{2}{\frac{\pi n}{N}}\quad(n=1,2,\ldots,N).\label{eq:gravitonspectrum}\end{equation}
 The spectrum in Eq.~(\ref{eq:gravitonspectrum}) describes one diagonal
zero--mode graviton which corresponds to the unbroken GC and a phonon--like
spectrum of massive gravitons that matches in the IR, \textit{i.e.},
in the regime $n\ll N$, onto a linear KK tower. At this level, the
phenomenology of the model appears to be very similar to that of a
deconstructed gauge theory. An important qualitative difference to
deconstruction, however, reveals itself in the peculiar strong coupling
effects of the theory.

It has been demonstrated in Ref.~\cite{Arkani-Hamed:2002sp}, that
the strong coupling behavior of discrete gravitational extra dimensions
is most conveniently exhibited by making use of the Callan--Coleman--Wess--Zumino
formalism for effective field theories \cite{Coleman:1969sm}. Following
this lead, the product symmetry group $\Pi_{i=1}^{N}GC_{i}$ can be
formally restored in $\mathcal{S}_{\textrm{link}}^{g}$ by appropriately
adding Goldstone bosons. To this end, one expands each link field
around the identity as $Y_{i}^{\mu}=x^{\mu}+\pi_{i}^{\mu}$, where
the Goldstone bosons $\pi_{i}^{\mu}$ transform non--linearly under
$GC_{i}$ and $GC_{i+1}$. The Goldstone bosons, which have three
polarizations, are eaten by the massless gravitons, which have two
polarizations, to generate the five polarizations of the massive gravitons
with spectrum as given in Eq.~(\ref{eq:gravitonspectrum}). Now,
the interactions of the lowest lying scalar longitudinal component
$\phi$ of the Goldstone bosons allow to extract directly the scale
of unitarity violation in the theory. It turns out that, for the model
at hand, the amplitude $\mathcal{A}(\phi\phi\rightarrow\phi\phi)$
for $\phi-\phi$ scattering is of the order $\mathcal{A}\sim E^{10}/\Lambda_{4}^{10}$,
where $E$ is the energy of $\phi$ and \begin{equation}
\Lambda_{4}=\left(\frac{M_{4}}{R^{3}}\right)^{1/4}\label{eq:Lambda4}\end{equation}
 is the strong coupling scale of the theory that is set by the triple
vertex of $\phi$. From Eq.~(\ref{eq:Lambda4}), it is seen that
the UV cutoff scale $\Lambda_{4}$ of the effective theory depends
on the IR length--scale $R$ of the compactified extra dimension.
This phenomenon has been called UV/IR connection \cite{Arkani-Hamed:2003vb}.
Since in a sensible effective theory for massive gravitons the lattice
spacing $m^{-1}$ must always be larger than the minimal lattice spacing
defined by $a_{\textrm{min}}\sim\Lambda_{4}^{-1}$, this implies that
the theory does not possess a naive continuum limit. In other words,
for given radius $R$, the effective theory is characterized by a
highest possible number of lattice sites $N_{\textrm{max}}=R\Lambda_{4}$,
which limits how fine grained the lattice can be made.

Besides the triple derivative coupling of $\phi$, the Goldstone boson
action contains other types of vertices, each of which can be associated
with a characteristic strong coupling scale for that interaction \cite{Schwartz:2003vj}.
As two such typical examples, we will consider the scales \begin{equation}
\Lambda_{3}=\left(\frac{M_{4}}{R^{2}}\right)^{1/3}\,\,\,\,\textrm{and}\,\,\,\,\Lambda_{5}=\left(\frac{M_{4}}{R^{4}}\right)^{1/5},\label{eq:Lambda35}\end{equation}
 which we will later compare with $\Lambda_{4}$. It is important
to note that the existence of the strong coupling scales in Eqs.~(\ref{eq:Lambda4})
and (\ref{eq:Lambda35}) is qualitatively different from the UV cutoff
in deconstructed gauge theories. In deconstruction, the strong coupling
scale associated with the non--linear sigma model approximation is always by a factor $\sim4\pi$
larger than the mass of the heaviest gauge boson, which is of the
order the inverse lattice spacing. In this sense, deconstruction may
provide, unlike the effective theory of massive gravitons discussed
here, an UV completion of higher--dimensional gauge theories. It should
be noted, however, that the emergence of the scales in Eqs.~(\ref{eq:Lambda4})
and (\ref{eq:Lambda35}) is a result of choosing a minimal discretization
with nearest--neighbor couplings and may be avoided in specific types
of non--local theory spaces \cite{Schwartz:2003vj}.

\section{Incorporation of matter}

\label{sec:latticizedmatter}

Let us now extend the model in Sec.~\ref{sec:gravity}, which has
been formulated for pure gravity, by adding on each site extra scalar
and fermionic site variables. To illustrate the general idea, we shall
restrict ourselves here, for simplicity, to the case where we have
on each site $i$ only one scalar $\Phi_{i}$ and one Dirac fermion
$\Psi_{i}$. We suppose that the sets of scalar and fermionic site
variables $\bigcup_{i=1}^{N}\Phi_{i}$ and $\bigcup_{i=1}^{N}\Psi_{i}$
respectively describe, in the sense of usual lattice gauge theory,
a scalar $\Phi$ and a fermion $\Psi$ propagating in the discretized
fifth dimension discussed in Sec.~\ref{sec:gravity}. The total action
$\mathcal{S}$ of our model can therefore be split into contributions
from the sites and links as \begin{equation}
\mathcal{S}=\sum_{X=g,\Phi,\Psi}(\mathcal{S}_{\textrm{site}}^{X}+\mathcal{S}_{\textrm{link}}^{X}),\label{eq:S}\end{equation}
 where we have distinguished between the purely gravitational part
($X=g$), which is given in Eqs.~(\ref{eq:gsite}) and (\ref{eq:FP}),
and the sum of contributions from the scalar ($X=\Phi$) and fermion
($X=\Psi$) species. Let us first specify in Eq.~(\ref{eq:S}) the
interactions $\mathcal{S}_{\textrm{site}}^{X}$ on the sites. For
the lattice fields $\Phi$ and $\Psi$ we take in Eq.~(\ref{eq:S})
the matter actions \begin{subequations}\label{eq:mattersites}
\begin{eqnarray}
\mathcal{S}_{\textrm{site}}^{\Phi} & = & \sum_{i=1}^N\int
\textrm{d}^{4}x\sqrt{g^{i}}
(-{\textstyle\frac{1}{2}})(g_{\mu\nu}^{i}\partial^{\mu}\Phi_{i}\partial^{\nu}\Phi_{i}+M_{\Phi}^{2}\Phi_i\Phi_{i}),\label{eq:scalarsites}\\
\mathcal{S}_{\textrm{site}}^{\Psi} & = & \sum_{i=1}^N\int\textrm{d}^{4}x\sqrt{g^{i}}[{\textrm{i}}(\overline{\Psi}_{i}\gamma^{\alpha}{V_{\alpha}}^{\mu}(\partial_{\mu}+\Gamma_{\mu})\Psi_{i}+M_{\Psi}\overline{\Psi}_{i}\Psi_{i}],\label{eq:fermionsites}\end{eqnarray}
\end{subequations} where $M_{\Phi}$ and $M_{\Psi}$ denote the bulk masses of the 5D
scalar $\Phi$ and fermion $\Psi$, respectively. In Eq.~(\ref{eq:fermionsites}),
we have written the fermion action using the vierbein formalism (see,
\textit{e.g.}, Ref.~\cite{vierbein}), where $\gamma^{\alpha}$ ($\alpha=0,1,2,3$)
are the usual Dirac gamma matrices, while ${V_{\alpha}}^{\mu}$ is
the vierbein and $\Gamma_{\mu}$ is the associated spin connection.
It is obvious, that the action $\sum_{X}\mathcal{S}_{\textrm{site}}^{X}$,
summarizing the interactions on the $N$ sites, is invariant under
$N$ copies of GC. The $N$-fold product of GC's $\Pi_{i=1}^{N}GC_{i}$,
however, is explicitly broken in Eq.~(\ref{eq:S}) by each term in
the sum $\sum_{X}\mathcal{S}_{\textrm{link}}^{X}$, which contains
the interactions between the fields on the different sites. On the
transverse lattice, we suppose that $\Phi$ and $\Psi$ are coupled
to their nearest neighbors via \begin{subequations}\label{eq:matterlinks}
\begin{eqnarray}
\mathcal{S}_{\textrm{link}}^{\Phi} & = & \sum_{i=1}^N\int\textrm{d}^{4}x\sqrt{g^{i}}m^2\Phi_{i}(\Phi_{i+1}-\Phi_{i})+{\textrm{h.c.}},\label{eq:scalarlink}\\
\mathcal{S}_{\textrm{link}}^{\Psi} & = & \sum_{i=1}^N\int\textrm{d}^{4}x\sqrt{g^{i}}
m\overline{\Psi}_{iL}(\Psi_{(i+1)R}-\Psi_{iR})+{\textrm{h.c.}},\label{eq:fermionlink}
\end{eqnarray}
\end{subequations} where $\Psi_{iL,R}=\frac{1}{2}(1\mp\gamma_{5})\Psi_{i}$, with $\gamma_{5}=\textrm{i}\gamma^{0}\gamma^{1}\gamma^{2}\gamma^{3}$,
are the left-- and right--handed components of the Dirac spinor $\Psi_{i}$.
To arrive at Eq.~(\ref{eq:fermionlink}), we started with the Wilson--Dirac
action \cite{Wilson:1974sk}\begin{eqnarray}
\mathcal{S}_{\textrm{W}} & = & \sum_{i=1}^{N}\int\textrm{d}^{4}x\sqrt{g^{i}}m\left(\overline{\Psi}_{i}\frac{r+\gamma_{5}}{2}\Psi_{i+1}+\overline{\Psi}_{i}\frac{r-\gamma_{5}}{2}\Psi_{i-1}-r\overline{\Psi}_{i}\Psi_{i}\right),\label{eq:Wilson}\end{eqnarray}
 where $r$ is some arbitrary parameter. The action in Eq.~(\ref{eq:Wilson})
results from adding a Wilson term (which would vanish in the continuum
limit $m\rightarrow\infty$) to the naive lattice action of fermions,
thereby projecting out unwanted fermion doublers. We then obtain from
$\mathcal{S}_{\textrm{W}}$ the action $\mathcal{S}_{\textrm{link}}^{\Psi}$
in Eq.~(\ref{eq:fermionlink}) by assuming for the parameter $r$
Wilson's choice $r=1$ \cite{Wilson:1977}. As a consequence, we arrive
at a common mass spectrum for scalars and fermions, which is given
by \begin{equation}
m_{n}^{2}=4m^{2}\sin^{2}{\frac{\pi n}{N}}+M_{X}^{2}\quad(n=1,2,\ldots,N),\label{eq:matterspectrum}\end{equation}
 where $X=\Phi,\Psi$. The assumption of Wilson--fermions as in Eq.~(\ref{eq:fermionlink})
with $r=1$ ensures for $M_{\Psi}=M_{\Phi}$ identical dispersion
relations for the latticized fermions and bosons. Notice also, that
Eq.~(\ref{eq:matterspectrum}) becomes for $X=g$ identical with
the graviton spectrum in Eq.~(\ref{eq:gravitonspectrum}), when setting
the bulk graviton mass to zero, \textit{i.e.}, $M_{g}=0$. In the
weak field limit, we observe that for even $N$, the action $\mathcal{S}_{\textrm{link}}^{\Psi}$
in Eq.~(\ref{eq:fermionlink}) is characterized by $N/2$ global
$Z_{2}$ symmetries%
\footnote{Discrete non--Abelian flavor symmetries from deconstruction have recently
been analyzed in Ref.~\cite{Kubo:2005ty}.%
}\begin{eqnarray}
Z_{2}^{(i)} & : & \Psi_{(i+k)L}\longrightarrow-\Psi_{(i-k)L},\quad\Psi_{(i+k)R}\longleftrightarrow\Psi_{(i-k+1)R},\quad h_{\mu\nu}^{i+k}\longleftrightarrow h_{\mu\nu}^{i-k},\label{eq:Z2}\end{eqnarray}
 where $i=1,2,\ldots,N/2$ is held fixed, while $k$ runs over all
the values $k=0,\pm1,\pm2,\ldots,\pm N/2$. Starting with the Wilson--Dirac
action $\mathcal{S}_{\textrm{W}}$ in Eq.~(\ref{eq:Wilson}), the
discrete symmetries $Z_{2}^{(i)}$ are only consistent with the form
of the action $\mathcal{S}_{\textrm{link}}^{\Psi}$ in Eq.~(\ref{eq:fermionlink}),
which is obtained for the choice $r=1$. We wish to point out, that
the {}``locality'' of the actions $\mathcal{S}_{\textrm{link}}^{X}$
with nearest neighbor couplings might be understood in terms of scale--invariant
renormalization group transformations acting in theory space \cite{Hill:2003cy}.

\section{Casimir energies}

\label{sec:matterenergy}In this section, we investigate the Casimir
energies of matter fields propagating in the discrete extra dimension
introduced in Secs.~\ref{sec:discretedimensions} and \ref{sec:latticizedmatter}.
For a continuous 5D space--time manifold, the Casimir energy densities
of free massless scalars and fermions have been computed in Ref.~\cite{Candelas:1983ae},
whereas the Casimir contribution of a massless graviton in the same
background, using the standard effective action theory, can be found
in Ref.~\cite{Appelquist:1982zs}. In our model with a discrete fifth
dimension, one can summarize in the 4D low--energy theory the vacuum
energy contributions of the massive modes to the 1--loop effective
potential as \begin{equation}
V_{\textrm{eff}}=(s-4f+5g)\sum_{n=1}^{N}V_{0}(m_{n}),\label{eq:Veff}\end{equation}
 where~$s,f,$ and~$g$ respectively denote the number of real scalar,
fermionic, and gravitational fields propagating in the latticized
bulk. In Eq.~(\ref{eq:Veff}), we have summed for each latticized
field over the vacuum energy densities~$V_{0}(m_{n})$ of all the
modes with masses~$m_{n}$ belonging to the phonon--like spectrum
in Eq.~(\ref{eq:matterspectrum}). Notice in Eq.~(\ref{eq:Veff}),
that the factors $-4$ and $5$ reflect the spin--degrees of freedom
that contribute to each massive fermion and graviton loop. In continuum
KK theories, a gauge--independent gravitational quantum--effective
action can be consistently formulated by employing the Vilkovisky--De
Witt effective action \cite{Vilkovisky:1984st}, for which, however,
only a few explicit examples in special topologies are known \cite{Cho:2000cb}.
For our model with a discrete extra dimension, the contribution $V_{0}(m_{n})$
to the effective potential from a single real scalar degree of freedom
with $M_{X}=0$ has been calculated in Refs.~\cite{Kan:2002rp,Cognola:2003xr},
where\[
V_{0}(m_{n})=\frac{m_{n}^{4}}{64\pi^{2}}\left(\ln\frac{m_{n}^{2}}{\mu^{2}}-\frac{3}{2}\right)\]
 has been obtained by a zeta--function regularization technique~\cite{Hawking:1976ja}.
In our theory space, the purely gravitational contribution to the
effective potential which includes only the tower of massive gravitons
{[}\textit{i.e.}, $s=f=0$ and $g=1$ in Eq.~(\ref{eq:Veff}){]},
for example, was then found to be\begin{equation}
\left.V_{\textrm{eff}}\right|_{s,f=0}=\frac{15Nm^{4}}{32\pi^{2}}\left(\textrm{ln}\frac{4m^{2}}{\mu^{2}}-\frac{3}{2}\right)+\frac{5m^{4}}{2\pi^{2}}\sum_{n=1}^{N-1}\sin^{4}\left(\frac{\pi n}{N}\right)\textrm{ln}\:\sin\left(\frac{\pi n}{N}\right),\label{eq:gravitoncontribution}\end{equation}
 where, from Eq.~(\ref{eq:gravitonspectrum}), $m_{n}^{2}=4m^{2}\sin^{2}{\frac{\pi n}{N}}$.
For a related discussion in a supersymmetric context see also Refs.~\cite{Nojiri:2004jm,Cognola:2005sc}.
Note that Eq.~(\ref{eq:gravitoncontribution}) contains also terms
that are not due to the Casimir effect or terms that depend on an
arbitrary renormalization scale~$\mu$ originating from the regularization
process.%
\footnote{The dependence on the renormalization scale~$\mu$ leads, in a cosmological
setup, to a running CC. Some recent work on such renormalization group
motivated DE models and their cosmological implications can be found
in Ref.~\cite{RGE-CC} and references therein.%
} Since we wish to consider only the 4D Casimir energy density, we
will, in the following, eliminate the unwanted parts in the effective
potential. This can be realized by subtracting off the vacuum energy
density that corresponds to an uncompactified (unbounded) extra dimension
as explained in Ref.~\cite{Bauer:2003mh}. As a nice advantage of
this renormalization procedure we obtain that the transverse lattice
result converges in the limit $N\rightarrow\infty$ \textit{exactly}
to the value expected from the continuum theory.

If the bulk masses $M_{X}$ of the fields in Eq.~(\ref{eq:Veff})
are all set to zero, the resulting 4D Casimir energy density of each
latticized bulk field would be of the order~$\sim R^{-4}$. As already
mentioned in the introduction, this would lead to the bound~$R\gtrsim0.1\,\textrm{mm}$.
Let us therefore now consider latticized matter fields with non--vanishing
bulk masses $M_{X}\neq0$. In the extra dimension, the boundary conditions
for the quantum fields can be periodic or anti--periodic, and the
corresponding fields are called untwisted and twisted, respectively.
The Casimir energy densities of these field configurations differ
by a factor of order one and have opposite sign. Following Ref.~\cite{Bauer:2003mh},
the 4D Casimir energy density of a single untwisted real scalar field
in the latticized fifth dimension can be written as\begin{equation}
\rho_{\textrm{untwisted}}=\frac{1}{2(2\pi)^{3}}\cdot\frac{4\pi}{8}\left[\sum_{n=1}^{N}m_{n}^{4}\ln m_{n}-N\cdot\int_{0}^{1}\textrm{d}s\cdot m_{s}^{4}\ln m_{s}\right],\label{eq:rho-lattice}\end{equation}
 where, from Eq.~(\ref{eq:matterspectrum}), $m_{n}^{2}=4m^{2}\sin^{2}(\pi n/N)+M_{X}^{2}$
and $s$ is treated in the integral as a continuous parameter which
replaces~$n/N$ in the sine function. As long as the number of lattice
sites is $N\gtrsim\mathcal{O}(10)$, the Casimir energy density on
the transverse lattice in Eq.~(\ref{eq:rho-lattice}) differs less
than $\lesssim1\%$ from the value in the naive continuum limit $N\rightarrow\infty$.
In the remainder of this section, we will therefore employ the expressions
for the Casimir energy densities of quantum fields in the continuum
theory. In this approximation, the vacuum energy density of a real
(un)twisted scalar field reads \cite{Bauer:2003mh}\begin{equation}
\rho_{\textrm{(un)twisted}}=\frac{\pm1}{8(2\pi)^{2}}\frac{(2\pi)^{5}}{R^{4}}\int_{x}^{\infty}\textrm{d}n\frac{(n^{2}-x^{2})^{2}}{\exp(2\pi n)\pm1},\label{eq:density}\end{equation}
 where the {}``$+$'' and {}``$-$'' signs belong to twisted and
untwisted fields, respectively, and $x=M_{X}R/(2\pi)$, in which $M_{X}$
denotes the bulk mass of the scalar field. The integral in Eq.~(\ref{eq:density})
can be performed exactly after neglecting the term~$\pm1$ in the
denominator, \textit{i.e.}, both densities differ only in an overall
sign:\begin{equation}
\rho_{\textrm{(un)twisted}}=\pm\frac{(M_{X}R)^{2}+3M_{X}R+3}{(2\pi)^{2}R^{4}}e^{-M_{X}R}.\label{eq:rho-(un)twisted}\end{equation}
 When taking the sum of contributions for twisted and untwisted fields,
the integrals must be added before carrying out the approximation,
which gives \begin{equation}
\rho_{\textrm{sum}}=-\frac{4(M_{X}R)^{2}+6M_{X}R+3}{16(2\pi)^{2}R^{4}}e^{-2M_{X}R}.\label{eq:rho-sum}\end{equation}
 The corresponding energy densities of Dirac fermions are obtained
by simply multiplying the scalar densities $\rho_{\textrm{(un)twisted}}$
by $-4$. Note that the applied approximation works fine even in the
limit of vanishing bulk masses $M_{X}\rightarrow0$. The basic feature
expressed in Eqs.~(\ref{eq:rho-(un)twisted}) and (\ref{eq:rho-sum})
is that for large bulk masses $M_{X}\gg R^{-1}$, the energy density
of massive matter fields becomes exponentially suppressed, which could compensate
for the possibly large factor $\sim R^{-4}$, even when $R$ is comparatively
small.

\begin{figure}
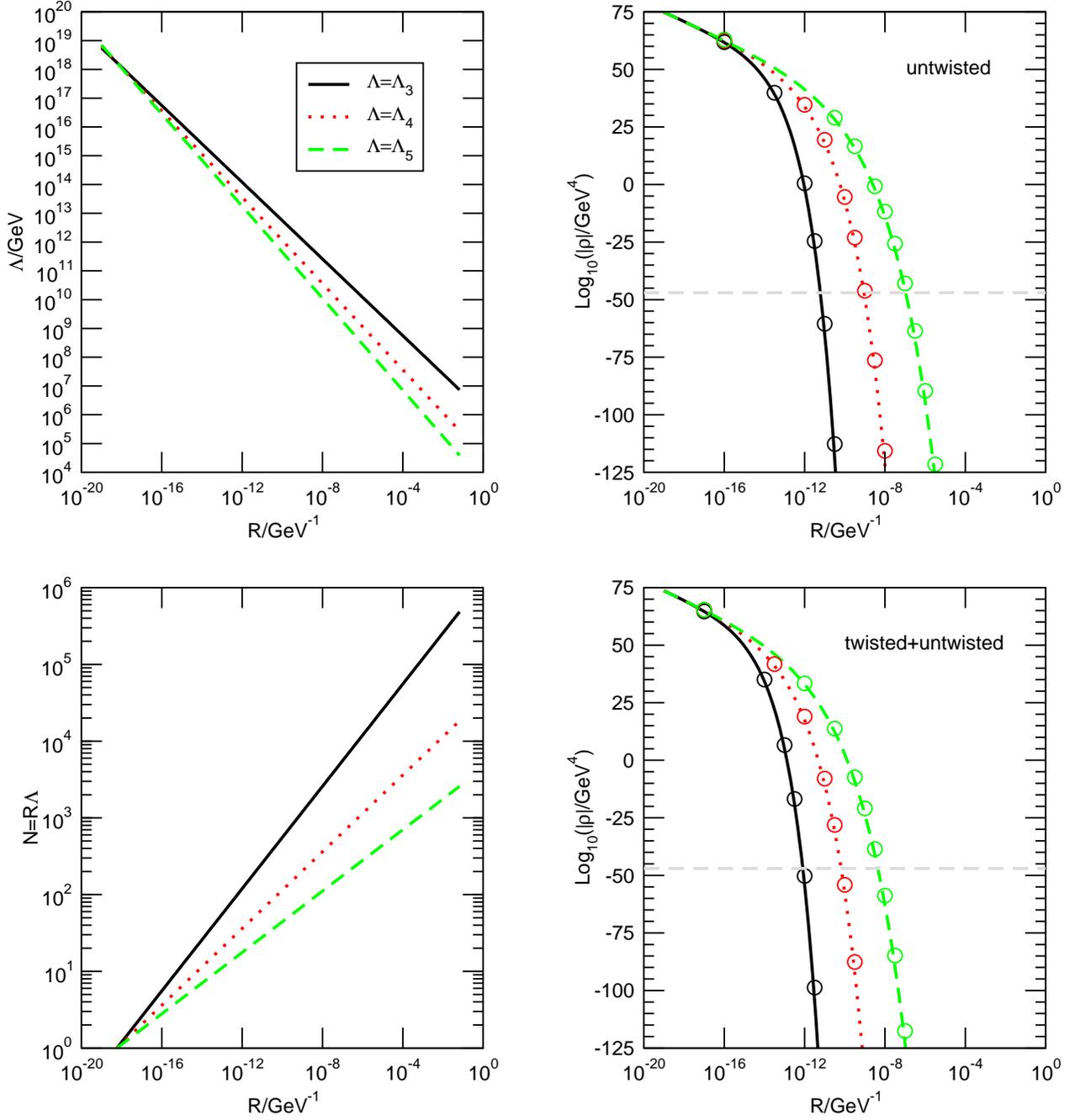

\begin{flushright}\includegraphics[%
  clip,
  scale=0.75]{UV-Casimir-Lambda-2.eps}\hspace{1cm}\includegraphics[%
  clip,
  scale=0.75]{UV-Casimir-rho-untwisted-mitGitter-2.eps}\vspace{0.5cm}\\
 \includegraphics[%
  clip,
  scale=0.75]{UV-Casimir-N-2.eps}\hspace{1cm}\includegraphics[%
  clip,
  scale=0.75]{UV-Casimir-rho-sum-mitGitter-2.eps}\end{flushright}

\caption{\label{fig:rho-Lambda-N}For the three choices~$\Lambda=\Lambda_{3},\Lambda_{4},\Lambda_{5}$
of the strong coupling scale~$\Lambda$ from Eqs.~(\ref{eq:Lambda4})
and~(\ref{eq:Lambda35}), we plotted the values of~$\Lambda$, the
Casimir energy densities~$\rho$, and the corresponding number~$N=R\Lambda$
of lattice sites as functions of the size~$R$ of the fifth dimension.
The energy densities~$\rho$ are given for the untwisted scalar field
{[}\textit{cf.}~Eq.~(\ref{eq:rho-(un)twisted}){]} and the sum of
one untwisted and one twisted scalar field {[}\textit{cf}.~Eq.~(\ref{eq:rho-sum}){]}.
Note, that~$\rho$ is negative in both cases, and the bulk masses
of the fields have their maximal values, given by~$\Lambda$, according
to Sec.~\ref{sec:matterenergy}. In the plots of~$\rho$, the horizontal
dashed line marks the observed value~$\rho_{\textrm{obs}}\sim10^{-47}\,\textrm{GeV}^{4}$
of the DE density and the circles represent exact lattice values from
Eq.~(\ref{eq:rho-lattice}).}
\end{figure}

Now, we are in a position to calculate the Casimir energy densities
with the bulk masses~$M_{X}$ set equal to the strong coupling scales
$\Lambda_{3}$, $\Lambda_{4}$, and $\Lambda_{5}$ given in Eqs.~(\ref{eq:Lambda4})
and~(\ref{eq:Lambda35}). The effective field theory description
suggests that these are the largest possible values that~$M_{X}$
can take in the gravitational theory space. If the UV cutoff $\Lambda$
is much larger than $\sim R^{-1}$, the expressions in Eqs.~(\ref{eq:rho-(un)twisted})
and (\ref{eq:rho-sum}) are dominated by the exponential damping factors,
such that the Casimir energy densities are most strongly suppressed
when $M_{X}$ becomes of the order the strong coupling scale $\Lambda$,
with $\Lambda=\Lambda_{3},\Lambda_{4},\Lambda_{5}$. Moreover, this
suppression is most effective, when the number of lattice sites $N$
is maximized by choosing the inverse lattice spacing~$m=N/R$ to
be also of the order $\Lambda$. The lower limit $R_{\textrm{min}}$
on the size~$R$ of the extra dimension emerges from requiring that
the Casimir energy densities remain below the observed value $\rho_{\textrm{obs}}\simeq10^{-47}\:\textrm{GeV}^{4}$
of the DE density. The results for an untwisted scalar field and the sum of twisted
and untwisted fields are plotted in Fig.~\ref{fig:rho-Lambda-N}.
Since the smallest value~$R_{\textrm{min}}$ that $R$ can take is,
due to the UV/IR connection, a function of $\Lambda$, we have considered~$R_{\textrm{min}}(\Lambda)$
for all three scales $\Lambda=\Lambda_{3},\Lambda_{4},\Lambda_{5}$.
These values together with the corresponding maximum number of lattice
sites ~$N=R_{\textrm{min}}\cdot\Lambda(R_{\textrm{min}})$, where
$\Lambda(R_{\textrm{min}})$ is the strong coupling scale associated
with $R_{\textrm{min}}$, are summarized in Tab.~\ref{tab:Rmin}.
Note that we can apply here the relations from the continuum theory,
since (i) the number of lattice sites~$N$ is of the order~$\sim10^{2}$
and (ii) the lattice calculation leads to energy densities (drawn
in Fig.~\ref{fig:rho-Lambda-N} as circles), that agree very well
with the values in the continuum theory.%
\footnote{For~$R_{\textrm{min}}$, the values of the continuum and lattice
formulas differ by about~$15\%$, which is negligible, since the
strong coupling scales~$\Lambda_{3,4,5}$ are order of magnitude
estimates. For instance, the lattice calculation for an untwisted
scalar field and~$\Lambda=\Lambda_{3}$ gives $R_{\textrm{min}}=6.8\cdot10^{-12}\,\textrm{GeV}^{-1}$,
whereas the continuum approximation yields~$R_{\textrm{min}}=6.1\cdot10^{-12}\,\textrm{GeV}^{-1}$.%
} For a mix of a twisted and an untwisted field, we observe that the
Casimir energy density of massive bulk fields exhibits a stronger
suppression due to the different signs of both components. From Fig.~\ref{fig:rho-Lambda-N},
we read off that the minimal radius $R_{\textrm{min}}$ of the discrete
gravitational extra dimension lies in the range \begin{equation}
(10^{12}\,\textrm{GeV})^{-1}\lesssim R_{\textrm{min}}\lesssim(10^{7}\,\textrm{GeV})^{-1},\label{eq:range}\end{equation}
 where, typically, $\Lambda(R_{\textrm{min}})\sim10^{2}\times R_{\textrm{min}}^{-1}$.
For a radius $R$ which is much smaller than the range given in Eq.~(\ref{eq:range}),
the Casimir energy densities of the bulk matter fields would significantly
exceed $\rho_{\textrm{obs}}$ and thus run into conflict with observation.
Of course, there may be other possible sources of DE which might be
responsible for the accelerated expansion of the universe, but it
seems unlikely that they could exactly cancel the potentially large
contributions from the Casimir effect in extra dimensions. %

\begin{table}
\begin{center}\begin{tabular}{|c|c|c|c|}
\hline 
untwisted&
 $R_{\textrm{min}}$&
 $\Lambda(R_{\textrm{min}})$&
 $N=R_{\textrm{min}}\cdot\Lambda(R_{\textrm{min}})$\tabularnewline
\hline
$\Lambda_{3}$&
 $6,1\cdot10^{-12}\,\textrm{GeV}^{-1}$&
 $3,6\cdot10^{13}\,\textrm{GeV}$&
 $219$\tabularnewline
\hline
$\Lambda_{4}$&
 $9,0\cdot10^{-10}\,\textrm{GeV}^{-1}$&
 $2,2\cdot10^{11}\,\textrm{GeV}$&
 $198$\tabularnewline
\hline
$\Lambda_{5}$&
 $1,1\cdot10^{-7}\,\textrm{GeV}^{-1}$&
 $1,7\cdot10^{9}\,\textrm{GeV}$&
 $179$ \tabularnewline
\hline
\end{tabular}\end{center}

\begin{center}\begin{tabular}{|c|c|c|c|}
\hline 
sum&
 $R_{\textrm{min}}$&
 $\Lambda(R_{\textrm{min}})$&
 $N=R_{\textrm{min}}\cdot\Lambda(R_{\textrm{min}})$\tabularnewline
\hline
$\Lambda_{3}$&
 $8,2\cdot10^{-13}\,\textrm{GeV}^{-1}$&
 $1,4\cdot10^{14}\,\textrm{GeV}$&
 $112$\tabularnewline
\hline
$\Lambda_{4}$&
 $6,6\cdot10^{-11}\,\textrm{GeV}^{-1}$&
 $1,6\cdot10^{12}\,\textrm{GeV}$&
 $103$\tabularnewline
\hline
$\Lambda_{5}$&
 $4,4\cdot10^{-9}\,\textrm{GeV}^{-1}$&
 $2,1\cdot10^{10}\,\textrm{GeV}$&
 $95$ \tabularnewline
\hline
\end{tabular}\end{center}

\caption{\label{tab:Rmin}Lower bound~$R_{\textrm{min}}$ on the size~$R$
of the extra dimension for an untwisted real scalar field and the
sum of a twisted and an untwisted scalar. Additionally, the values
of the strong coupling scale~$\Lambda$ and the number of lattice
sites $N$ are given when~$R$ is equal to~$R_{\textrm{min}}$.
For the scale~$\Lambda$, we considered each of the three choices~$\Lambda=\Lambda_{3},\Lambda_{4},\Lambda_{5}$
from Eqs.~(\ref{eq:Lambda4}) and~(\ref{eq:Lambda35}). The lower
bound~$R_{\textrm{min}}$ emerges from the requirement that the absolute
Casimir energy density lies below the observed value $\rho_{\textrm{obs}}$
of the DE density, when the bulk field mass $M_{X}$ in Eq.~(\ref{eq:matterspectrum})
takes the largest possible value $M_{X}\simeq\Lambda$.}
\end{table}

Let us now briefly comment on the gravitational contribution to the
1--loop quantum effective action~$V_{\textrm{eff}}$ in Eq.~(\ref{eq:Veff}).
For zero bulk mass~$M_{g}=0$, the gravitational effective potential
given in Eq.~(\ref{eq:gravitoncontribution}) would lead to a contribution
to~$V_{\textrm{eff}}$ of the order~$\sim m^{4}$. The gravitational
vacuum energy, however, can be canceled in our model at the linear
level, when we assume the presence of a suitable number of latticized
matter fields with actions as given in Eqs.~(\ref{eq:mattersites})
and~(\ref{eq:matterlinks}), which have vanishing bulk masses~$M_{X}=0$.
For instance, choosing~$b=3$ massless scalars and~$f=2$ massless
fermions, we find from Eq.~(\ref{eq:Veff}) that in this case~$V_{\textrm{eff}}=0$,
which holds in linearized gravity at the 1--loop level for an arbitrary
number~$N$ of lattice sites. In this approximation, the cancelation
of bosonic and fermionic degrees of freedom would actually be approached
in the limit~$N\rightarrow\infty$ for any value of the parameter
$r$ in the Wilson--Dirac action in Eq.~(\ref{eq:Wilson}). The requirement
that this cancelation holds for arbitrary, \textit{i.e.}, also for
small $N$, however, uniquely singles out Wilsons's choice~$r=1$.
It is interesting to consider a possible origin of free massless scalars
in effective field theories for KK modes. In a $4+d$ dimensional
KK theory with~$d=4$ compactified extra dimensions, \textit{e.g.},
we would have in the 4D low--energy theory one tower of massive spin--2
states, three towers of massive spin--1 states and six towers of massive
spin--0 states with degenerate masses (see, \textit{e.g.}, Ref.~\cite{Han:1998sg}).
The effective potential of these fields could, in a similar way as
mentioned above, be canceled at the 1--loop level by adding only free
Dirac fermions with zero bulk masses. Notice that, since the massless
fields couple only gravitationally to the visible sector, a sufficiently
low temperature of the massless states would allow to retain the predictions
of standard big bang nucleosynthesis \cite{Chacko:2004cz}. Finally,
we note that the cancelation of vacuum energies in a supersymmetric
multi--graviton theory on space--times with non--trivial topology
was also considered very recently in Ref.~\cite{Cognola:2005sc},
where bulk masses and different boundary conditions were taken into
account.

\section{\label{sec:Conclusions}Summary and conclusions}

In this paper, we have analyzed the Casimir effect of matter fields
in the background of an effective 5D space--time. The underlying model
of a discrete gravitational extra dimension exhibits a strong coupling
behavior at an energy scale $\Lambda$, which depends via an UV/IR
connection non--trivially on the size $R$ of the extra dimension.
For a small compactified extra dimension, massless quantum fields
usually lead, due to the Casimir effect, to large vacuum energy contributions,
which are in stark contrast to current observations. To circumvent
this problem, we have assumed for the matter fields large bulk masses
$M_{X}$ to suppress the Casimir energy density exponentially, even
for a tiny extra dimension. However, the strong coupling scale sets
an upper bound on the values of the bulk masses $M_{X}\lesssim\Lambda$,
and therefore limits the suppression effect. This yields a lower bound
on the size of the fifth dimension, when the bulk masses take the maximal possible value $M_{X}\simeq\Lambda$. Here, we found that
the minimal size $R_{\textrm{min}}$ of the extra dimension lies in
the range~$R_{\textrm{min}}\sim(10^{12}\,\textrm{GeV})^{-1}\dots(10^{7}\,\textrm{GeV})^{-1}$
and that the corresponding maximum number of lattice sites is of the
order $\sim10^{2}$. Furthermore, we discussed the possibility of
canceling the contribution of massless bulk fields to the quantum
effective potential. Generally, it would be interesting to explore a possible
relation of our model to holography, as suggested by the UV/IR connection
\cite{Jejjala:2003qg}, and analyze also supersymmetric realizations
\cite{Nojiri:2004jm,Cognola:2005sc}, {\it e.g.}, in the framework of
sequestered sector models of anomaly mediation \cite{Gregoire:2004ic}.

\section*{Acknowledgments}

We would like to thank T.~Enkhbat for useful comments and discussions.
This work was supported by the {}``Sonderforschungsbereich 375 für
Astroteilchenphysik der Deutschen Forschungsgemeinschaft'' (F.B.)
and the U.S. Department of Energy under Grant Numbers DE-FG02-04ER46140
and DE-FG02-04ER41306 (G.S.). F.B. wishes to thank the Freistaat Bayern
for a {}``Promotionsstipendium''.

\end{document}